 \documentclass[pmlr,twocolumn]{jmlr} 
 \setcitestyle{square}
 \setcitestyle{comma}

\usepackage{booktabs}
\usepackage{paralist}
\usepackage{caption}
\usepackage[load-configurations=version-1]{siunitx} 

\title[CheXphoto]{CheXphoto: 10,000+ Photos and Transformations of Chest X-rays for Benchmarking Deep Learning Robustness}

\hypersetup{linkcolor=black}
\author{ 
 \Name{Nick {A. Phillips}}\nametag{\thanks{These authors contributed equally to this work}} \Email{nphill22@stanford.edu}\\
 \Name{Pranav Rajpurkar}\nametag{\footnotemark[1]} \Email{pranavsr@stanford.edu}\\
 \Name{Mark Sabini}\nametag{\footnotemark[1]} \Email{msabini@alumni.stanford.edu}\\
 \Name{Rayan Krishnan} \Email{rayank@stanford.edu}\\
 \Name{Sharon Zhou} \Email{sharonz@cs.stanford.edu}\\
 \Name{Anuj Pareek} \Email{anujpare@stanford.edu}\\
 \Name{Nguyet {Minh Phu}} \Email{minhphu@stanford.edu}\\
 \Name{Chris Wang} \Email{chrwang@stanford.edu}\\
 \Name{Mudit Jain} \Email{muditjai@gmail.com}\\
 \Name{Nguyen {Duong Du}} \Email{v.dunguyen@vinbrain.net}\\
 \Name{Steven {QH Truong}} \Email{Brain01@vinbrain.net}\\
 \Name{Andrew {Y. Ng}} \Email{ang@cs.stanford.edu}\\
 \Name{Matthew {P. Lungren}} \Email{mlungren@stanford.edu}\\
}

\begin{document}

\maketitle

\begin{abstract}
Clinical deployment of deep learning algorithms for chest x-ray interpretation requires a solution that can integrate into the vast spectrum of clinical workflows across the world. An appealing approach to scaled deployment is to leverage the ubiquity of smartphones by capturing photos of x-rays to share with clinicians using messaging services like WhatsApp. However, the application of chest x-ray algorithms to photos of chest x-rays requires reliable classification in the presence of artifacts not typically encountered in digital x-rays used to train machine learning models. We introduce CheXphoto, a dataset of smartphone photos and synthetic photographic transformations of chest x-rays sampled from the CheXpert dataset. To generate CheXphoto we (1) automatically and manually captured photos of digital x-rays under different settings, and (2) generated synthetic transformations of digital x-rays targeted to make them look like photos of digital x-rays and x-ray films. We release this dataset as a resource for testing and improving the robustness of deep learning algorithms for automated chest x-ray interpretation on smartphone photos of chest x-rays.
\end{abstract}

\begin{figure*}[htbp]
    \floatconts
    {fig:chexphoto}
    {\caption{Overview of the CheXphoto data generation process.}}
    {\includegraphics[width=\linewidth]{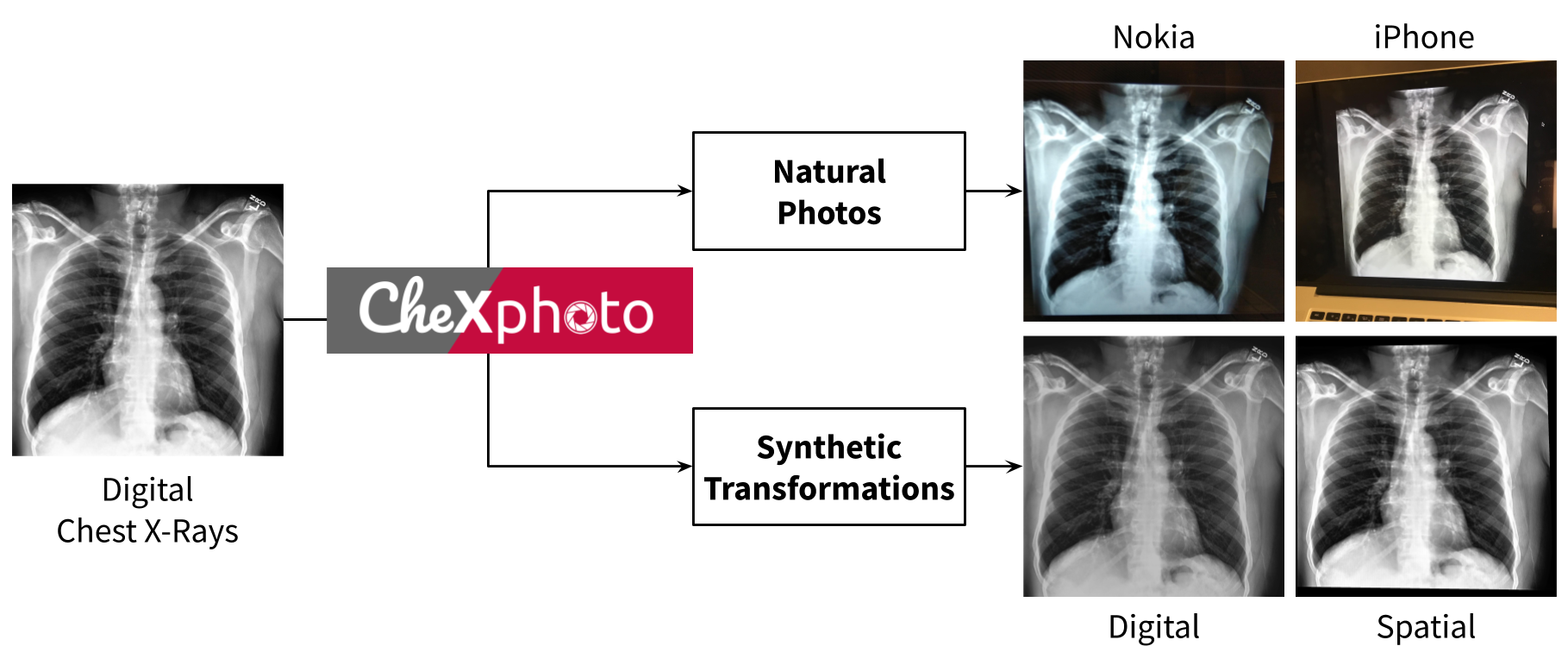}}
\end{figure*}

\section{Background \& Summary}
\label{sec:background}

Chest x-rays are the most common imaging exams, critical for diagnosis and management of many diseases and medical procedures. With over 2 billion chest x-rays performed globally each year, many clinics in both developing and developed countries have an insufficient number of radiologists to perform timely x-ray interpretation \citep{withey2019, Andronikou2011}. Computer algorithms could help reduce the shortage for x-ray interpretation worldwide.

Several recent advances in training deep learning algorithms for automated chest x-ray interpretation have been made possible by large datasets \citep{Nam2019, performance2}. In controlled settings, these deep learning algorithms can learn from labeled data to automatically detect pathologies at an accuracy comparable to that of practicing radiologists \citep{kallianos_ai_radiography}. These developments have been fueled by both improvements in deep learning algorithms for image classification tasks \citep{Huang2016, chexnext}, and by the release of large public datasets \citep{chexpert, chestxray8, bustos2019padchest, johnson2019mimiccxrjpg}. Although these algorithms have demonstrated the potential to provide accurate chest x-ray interpretation and increase access to radiology expertise, major obstacles remain in their translation to the clinical setting \citep{performance1, performance2}.

One significant obstacle to the adoption of chest x-ray algorithms is that deployment requires a solution that can integrate into the vast spectrum of clinical workflows around the world. Most chest x-ray algorithms are developed and validated on digital x-rays, while the majority of developing regions use films \citep{mobile_teleradiology, Andronikou2011}. An appealing approach to scaled deployment is to leverage the ubiquity of existing smartphones: automated interpretation of x-ray film through cell phone photography has emerged through a ``store-and-forward telemedicine'' approach, in which one or more digital photos of chest films are sent as email attachments or instant messages by practitioners to obtain second opinions from specialists as part of clinical care \citep{xray_cell_phone, Vassallo125}. Furthermore, studies have shown that photographs of films using modern phone cameras are of equivalent diagnostic quality to the films themselves \citep{mobile_teleradiology}, indicating the feasibility of high-quality automated algorithmic interpretation of photos of x-ray films.

Automated interpretation of chest x-ray photos at the same level of performance as digital chest x-rays is challenging because photography introduces visual artifacts not commonly found in digital x-rays, such as altered viewing angles, variable ambient and background lighting conditions, glare, moir\'e, rotations, translations, and blur \citep{adversarial_examples_world}. Image classification algorithms have been shown to experience a significant drop in performance when input images are perceived through a camera \citep{adversarial_examples_world}. Although recent work has demonstrated good generalizability of deep learning algorithms trained on digital x-rays to photographs \citep{chexpedition}, interpretation performance could be improved through inclusion of x-ray photography in the training process \citep{hendrycks_robustness, adversarial_explaining}. However, there are currently no large-scale public datasets of photos of chest x-rays.

To meet this need, we developed CheXphoto, a dataset of photos of chest x-rays and synthetic transformations designed to mimic the effects of photography. We believe that CheXphoto will enable researchers to improve and evaluate model performance on photos of x-rays, reducing the barrier to clinical deployment.

\section{Methods}
We introduce CheXphoto, a dataset of photos of chest x-rays and synthetic transformations designed to mimic the effects of photography. Specifically, CheXphoto includes a set of (1) \textit{Natural Photos}: automatically and manually captured photos of x-rays under different settings, including various lighting conditions and locations, and (2) \textit{Synthetic Transformations}: targeted transformations of digital x-rays to simulate the appearance of photos of digital x-rays and x-ray films. The x-rays used in CheXphoto are primarily sampled from CheXpert, a large dataset of 224,316 chest x-rays of 65,240 patients, with associated labels for 14 observations from radiology reports \citep{chexpert}.

CheXphoto comprises a training set of natural photos and synthetic transformations of 10,507 x-rays from 3,000 unique patients that were sampled at random from the CheXpert training set, and validation and test sets of natural and synthetic transformations of all 234 x-rays from 200 patients and 668 x-rays from 500 patients in the CheXpert validation and test sets, respectively. In addition, the CheXphoto validation set includes 200 natural photos of physical x-ray films sampled from external data sources, intended to more closely simulate pictures taken by radiologists in developing world clinical settings. As much of the developing world performs x-ray interpretation on film, this distinct set of images enables users to perform additional validation on a novel task that may be encountered in clinical deployment.

\subsection{Acquiring Natural Photos of Chest X-Rays}
Natural photos consist of x-ray photography using cell phone cameras in various lighting conditions and environments. We developed two sets of natural photos: images captured through an automated process using a Nokia 6.1 cell phone, and images captured manually with an iPhone 8. 

\begin{figure*}[!t]
    \floatconts
    {fig:chexpeditor}
    {\caption{Acquiring Natural Photos of Chest X-Rays Using Automated Capture \textbf{a}. Visual representation of the automated picture-taking process used for Nokia10k. The steps are described: 1. X-ray retrieved from computer storage, 2. X-ray displayed on monitor, 3. X-ray index and metadata sent to phone over UDP, 4. Index verified by phone, and camera triggered, 5. Application UI updated with new picture and filename, 6. Picture saved to phone storage with metadata in filename, 7. Computer notified that imaging was successful. 
    \textbf{b}. The physical setup used for Nokia10k, set in an example environment.
    \textbf{c}. Phone application UI, displaying most recent picture and saved filename.}}
    {\includegraphics[width=\textwidth]{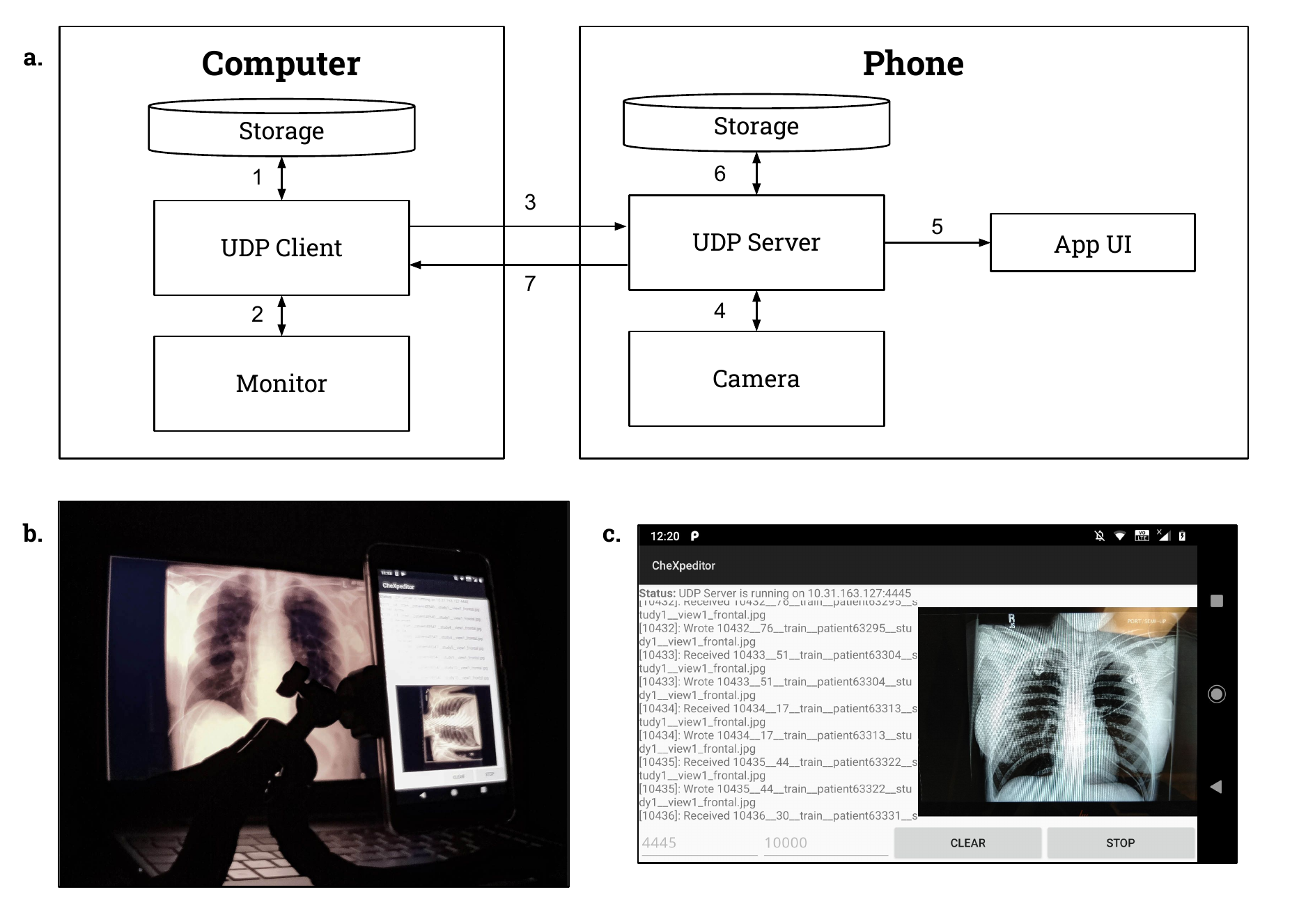}}
\end{figure*}

\subsubsection{Automated Capture of Nokia10k dataset}
We developed the `Nokia10k' dataset by capturing 10,507 images of digital chest x-rays using a tripod-mounted Nokia 6.1 cell phone (16 megapixel camera with a Zeiss sensor) and a custom Android application termed CheXpeditor to fully automate the processes of photography and metadata management. The primary challenge in automation was synchronizing picture-taking on the phone with displaying the chest x-ray on the monitor, to maintain a 1-to-1 correspondence between each chest x-ray and its photographed image. Without a robust synchronization method, photos of chest x-rays might be skipped or duplicated, jeopardizing the data collection process. Thus, bidirectional communication over UDP was established between the phone and the computer driving the monitor to exchange image metadata, take photos, and advance the chest x-ray on the monitor.

\begin{table*}[htbp]
\floatconts
{tab:label-counts}
{\caption{ The distribution of labeled observations for the Nokia10k training dataset.}}
{\begin{tabular}{llll}
                  Pathology &  Positive (\%) & Uncertain (\%) &   Negative (\%) \\
\hline
                 No Finding &    972 (9.25) &       0 (0.00) &   9535 (90.75) \\
 Enlarged Cardiomediastinum &    518 (4.93) &    600 (5.71) &   9389 (89.36) \\
               Cardiomegaly &   1313 (12.50) &    370 (3.52) &   8824 (83.98) \\
               Lung Opacity &  5184 (49.34) &    213 (2.03) &   5110 (48.63) \\
                Lung Lesion &    415 (3.95) &     78 (0.74) &  10014 (95.31) \\
                      Edema &   2553 (24.30) &    634 (6.03) &   7320 (69.67) \\
              Consolidation &    671 (6.39) &  1315 (12.52) &    8521 (81.10) \\
                  Pneumonia &     263 (2.50) &    885 (8.42) &   9359 (89.07) \\
                Atelectasis &  1577 (15.01) &  1595 (15.18) &   7335 (69.81) \\
               Pneumothorax &    957 (9.11) &    166 (1.58) &   9384 (89.31) \\
           Pleural Effusion &  4115 (39.16) &    607 (5.78) &   5785 (55.06) \\
              Pleural Other &    170 (1.62) &    127 (1.21) &  10210 (97.17) \\
                   Fracture &    391 (3.72) &      31 (0.30) &  10085 (95.98) \\
            Support Devices &  5591 (53.21) &     48 (0.46) &   4868 (46.33) \\
\hline
\end{tabular}}
\end{table*}

The 10,507 x-rays in Nokia10k were indexed deterministically from $1$ to $N$. We selected disjoint subsets of 250 to 500 consecutive indices to be photographed in constant environmental conditions. For each subset of indices, photography was conducted as follows:

\begin{inparaenum}[1)]
\item The $i$th chest x-ray was retrieved from computer storage.
\item The $i$th chest x-ray was displayed on the monitor.
\item The image metadata $m$ was assembled by the computer, and $(i, m)$ were sent to the phone via UDP.
\item The phone verified that $i$ was one greater than the previous index. If so, its camera was triggered. Else, the computer was notified of an error, and the entire picture-taking process was aborted.
\item The phone application UI, responsible for displaying status and current image, was updated to show the new picture and filename.
\item The picture was saved to phone storage with the metadata $m$ embedded in the filename.
\item The phone notified the computer that the imaging was successful, and the entire process was repeated for the $i+1$st chest x-ray.
\end{inparaenum}

After all images for a Nokia10k subset were taken, they were exported in one batch from the phone to storage. The metadata was parsed from the image filenames and used to automatically assign the correct CheXpert label. Alterations made to the imaging conditions after every subset included moving to a different room, switching the room light on/off, opening/closing the window-blinds, rotating the phone orientation between portrait/landscape, adjusting the position of the tripod, moving the mouse cursor, varying the monitor's color temperature, and switching the monitor's screen finish between matte/glossy. In all conditions, the chest x-ray was centered in the camera view-finder and lung fields were contained within the field of view.

\subsubsection{Manual Capture of iPhone1k dataset}
We developed the `iPhone1k dataset' by manually capturing 1,000 images of digital chest x-rays using an iPhone 8 (12 megapixel camera with a Sony Exmor RS sensor). The digital x-rays selected for the iPhone1k dataset are a randomly sampled subset of the x-rays used in the Nokia10k dataset. To produce the iPhone1k dataset, chest x-rays were displayed in full-screen on a computer monitor with 1920 x 1080 screen resolution and a black background. A physician took photos of the chest x-rays with a handheld iPhone 8 using the standard camera app. The physician was advised to change angle and distance from the computer monitor in-between each picture within constraints. 

For all images, the chest x-ray was centered in the viewfinder of the camera, and the thoracic and lung fields were contained within the field of view. Conformant to radiological chest x-ray standards, both lung-apices and costodiaphragmatic recesses were included craniocaudally, and the edges of the ribcage were included laterally. Photos were captured in sets of 100 to 200 images at a time; between sets, ambient alterations were made, such as switching the room-lighting on/off, opening or closing of the window-blinds, and physically moving the computer monitor to a different location in the room.

\subsection{Generating Synthetic Photographic Transformations of Chest X-Rays}

Synthetic transformations consist of automated changes to digital x-rays designed to simulate the appearance of photos of digital x-rays and x-ray films. We developed two sets of complementary synthetic transformations: digital transformations to alter contrast and brightness, and spatial transformations to add glare, moir\'e effects and perspective changes. To ensure that the level of these transformations did not impact the quality of the image for physician diagnosis, the images were verified by a physician. In some cases, the effects may be visually imperceptible, but may still be adversarial for classification models. For both sets, we apply the transformations to the same 10,507 digital x-rays selected for the Nokia10k dataset.

\begin{table*}[!t]
\floatconts
{tab:transforms}
{\caption{\label{tab:examples} Natural Photos (\textbf{a-b}) and Synthetic Transformations (Digital (\textbf{c-f}) and Spatial (\textbf{g-i})) included in CheXphoto.}}
{\begin{tabular}{cc}
  \includegraphics[width=30mm]{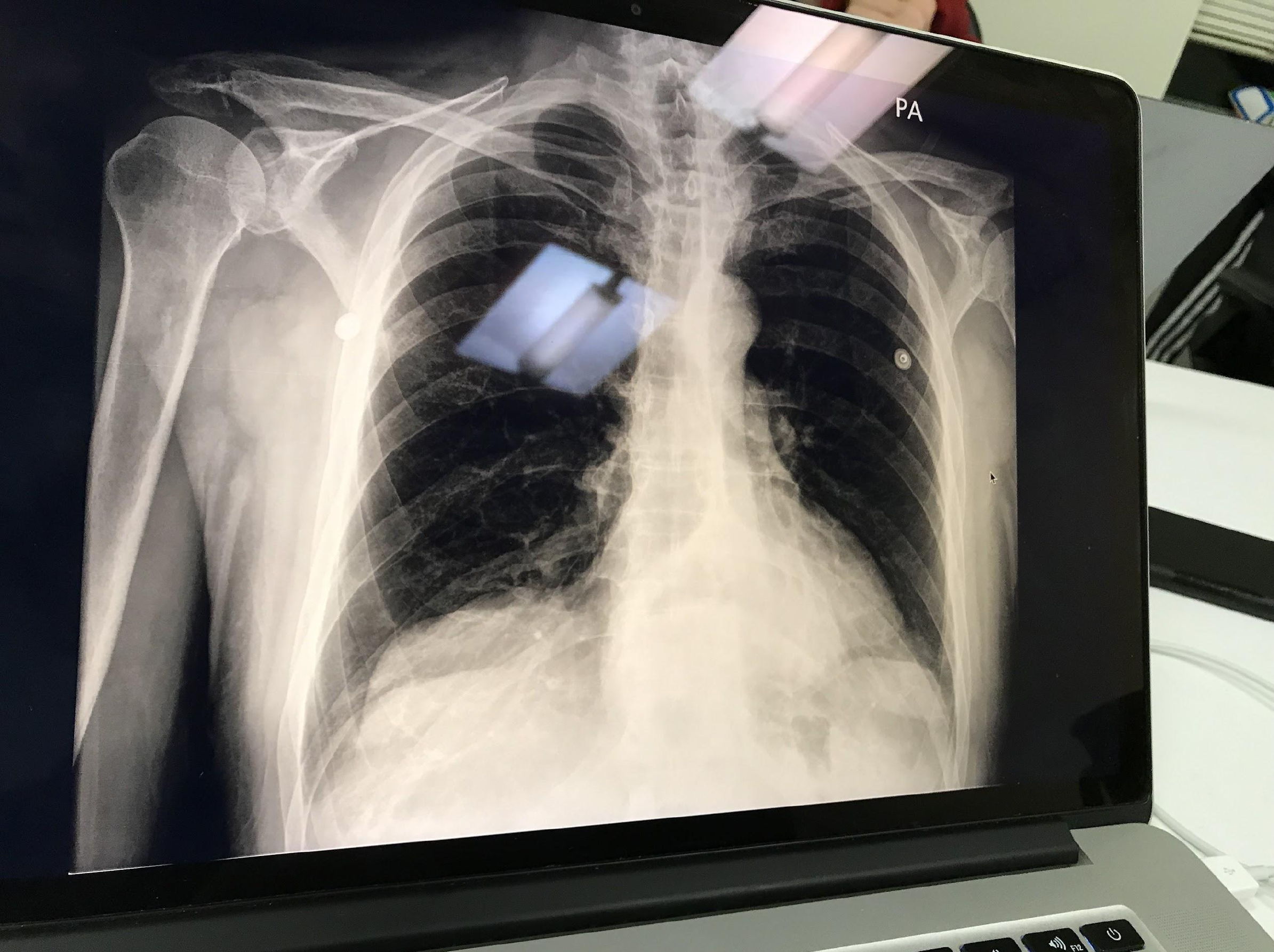} &   
  \includegraphics[width=30mm]{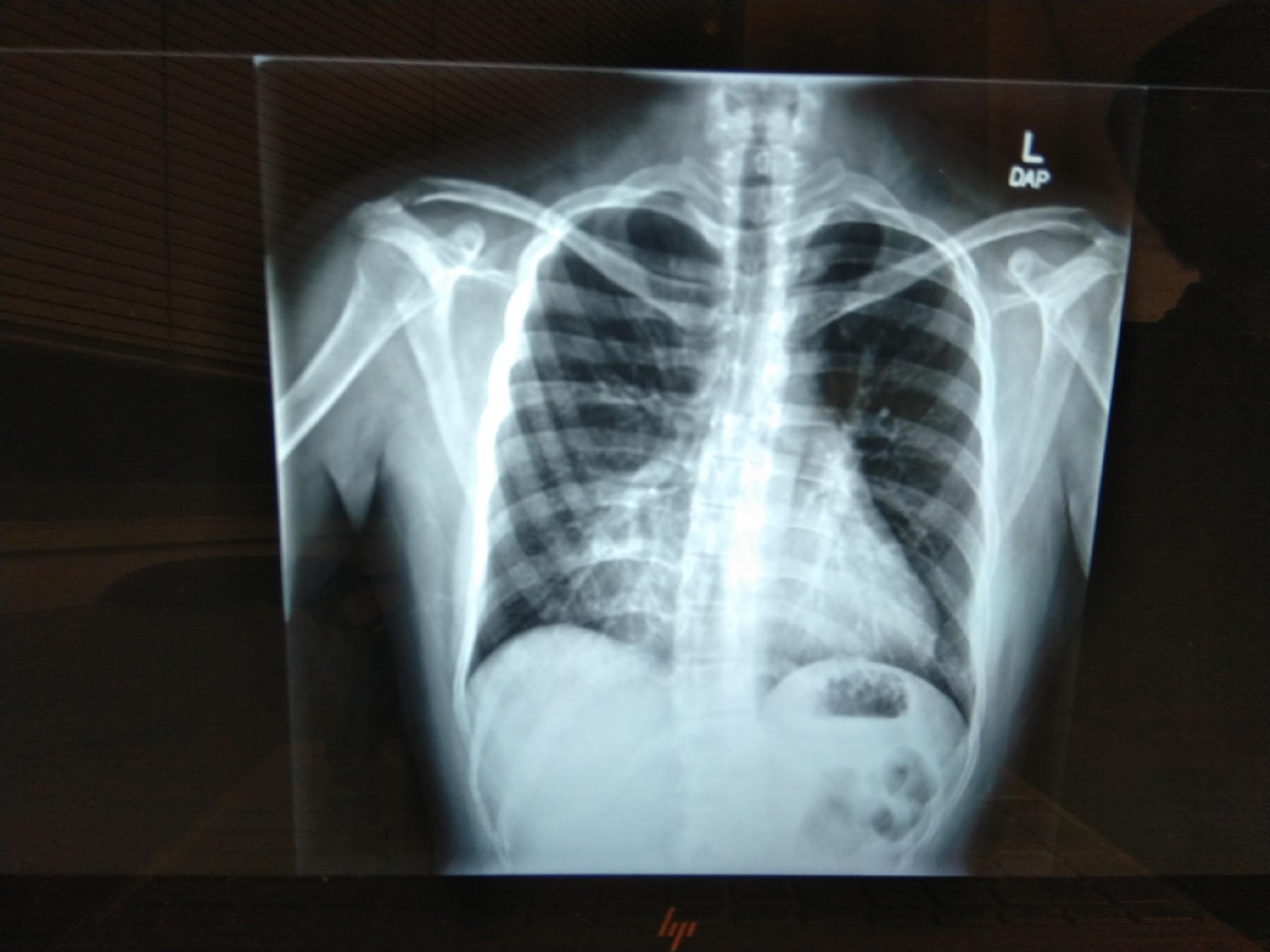} \\
(a) iPhone & (b) Nokia \\[6pt]
\end{tabular}
\begin{tabular}{cccc}
 \includegraphics[width=30mm]{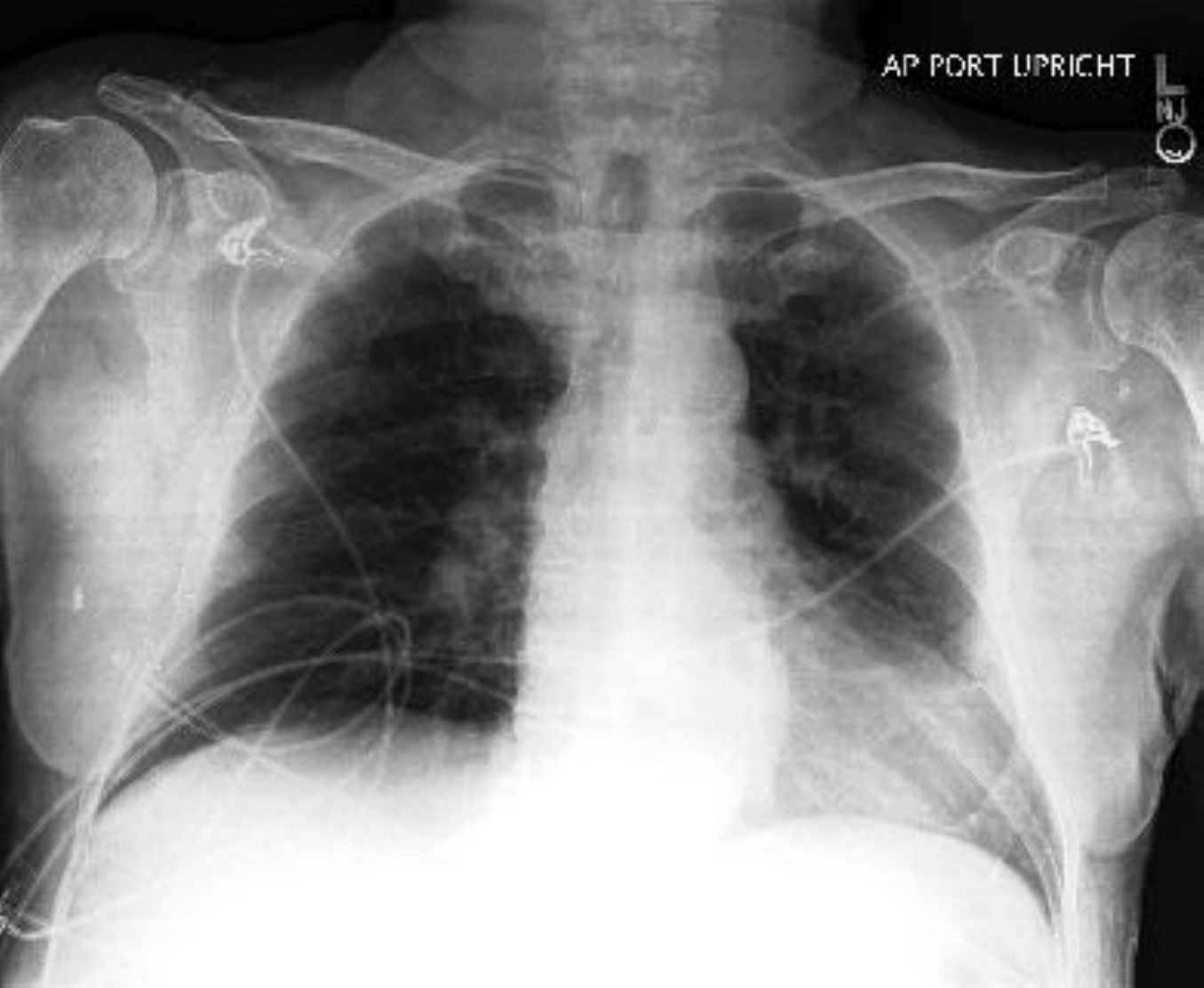} &   \includegraphics[width=30mm]{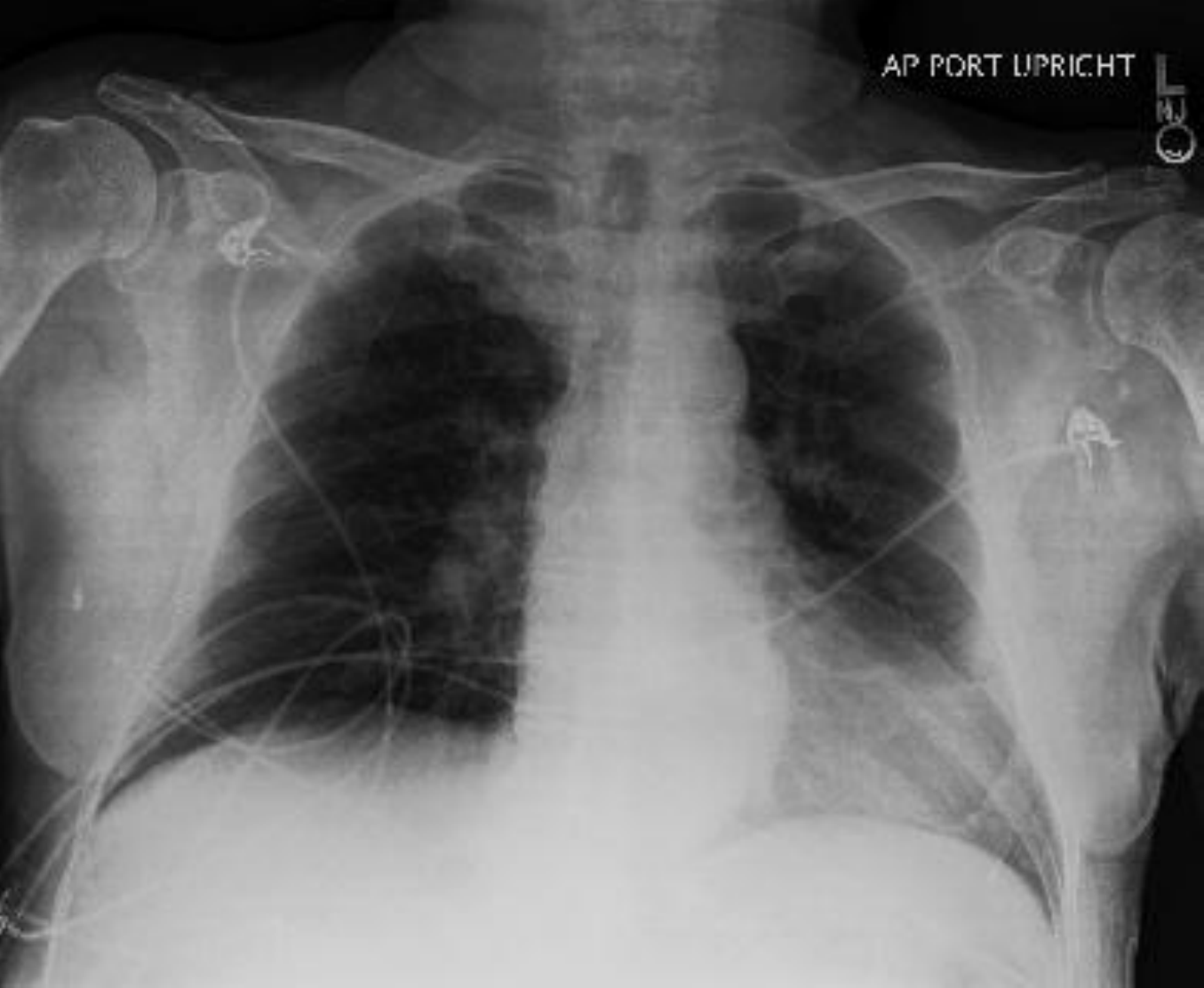} &
 \includegraphics[width=30mm]{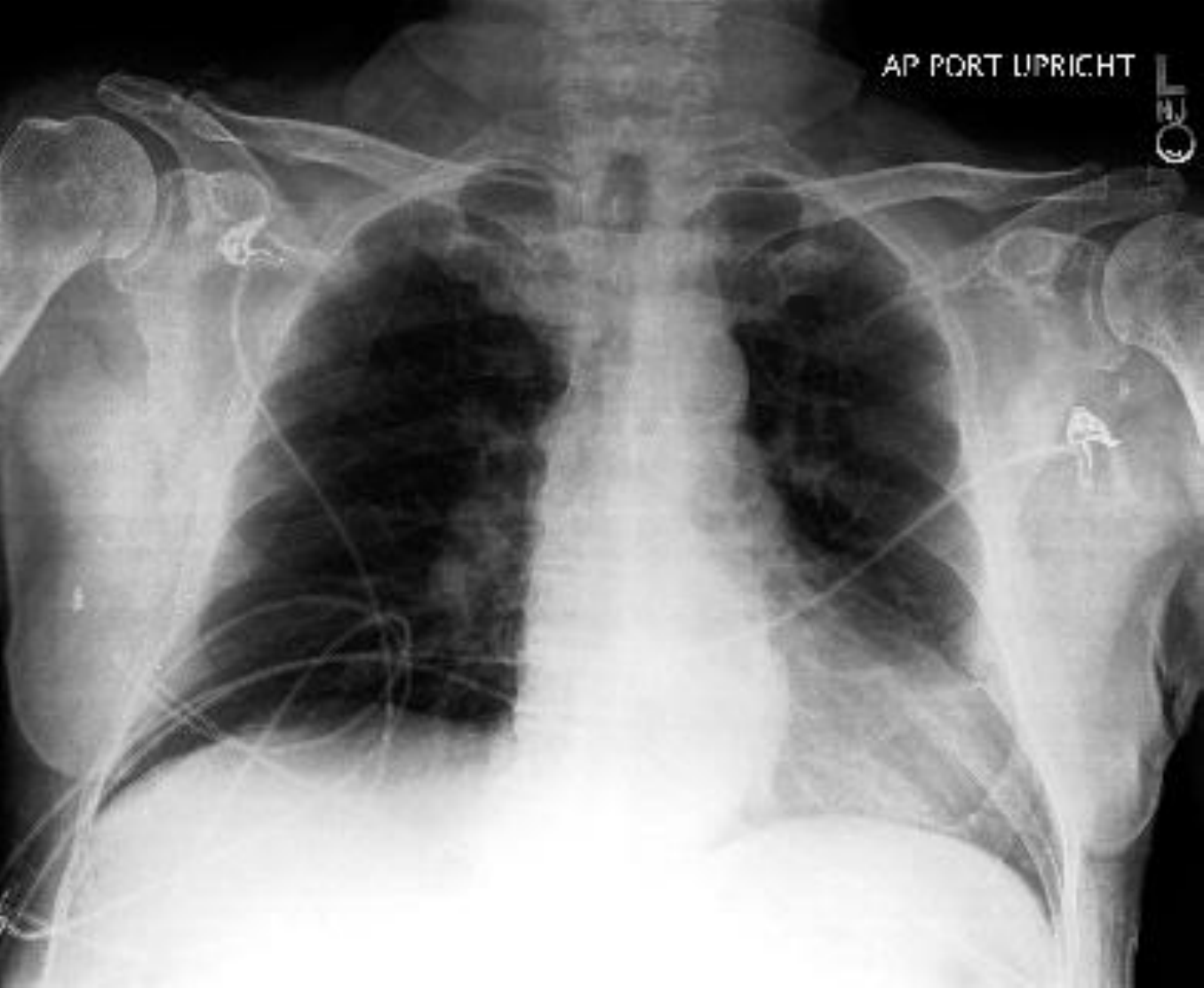} &
 \includegraphics[width=30mm]{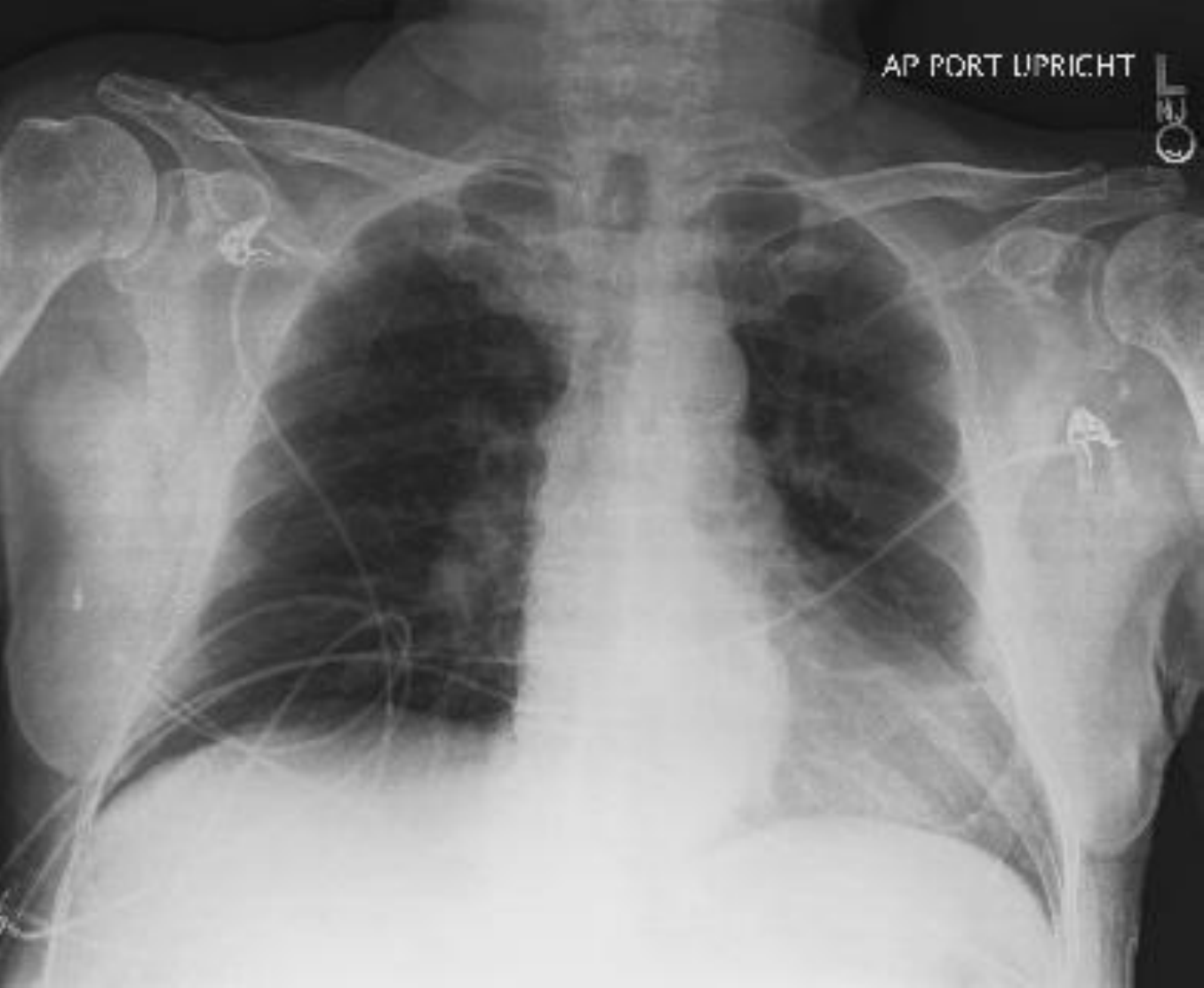} \\
(c) Brightness Up & (d) Brightness Down & (e) Contrast Up & (f) Contrast Down \\[6pt]
\end{tabular}
\begin{tabular}{ccc}
  \includegraphics[width=30mm]{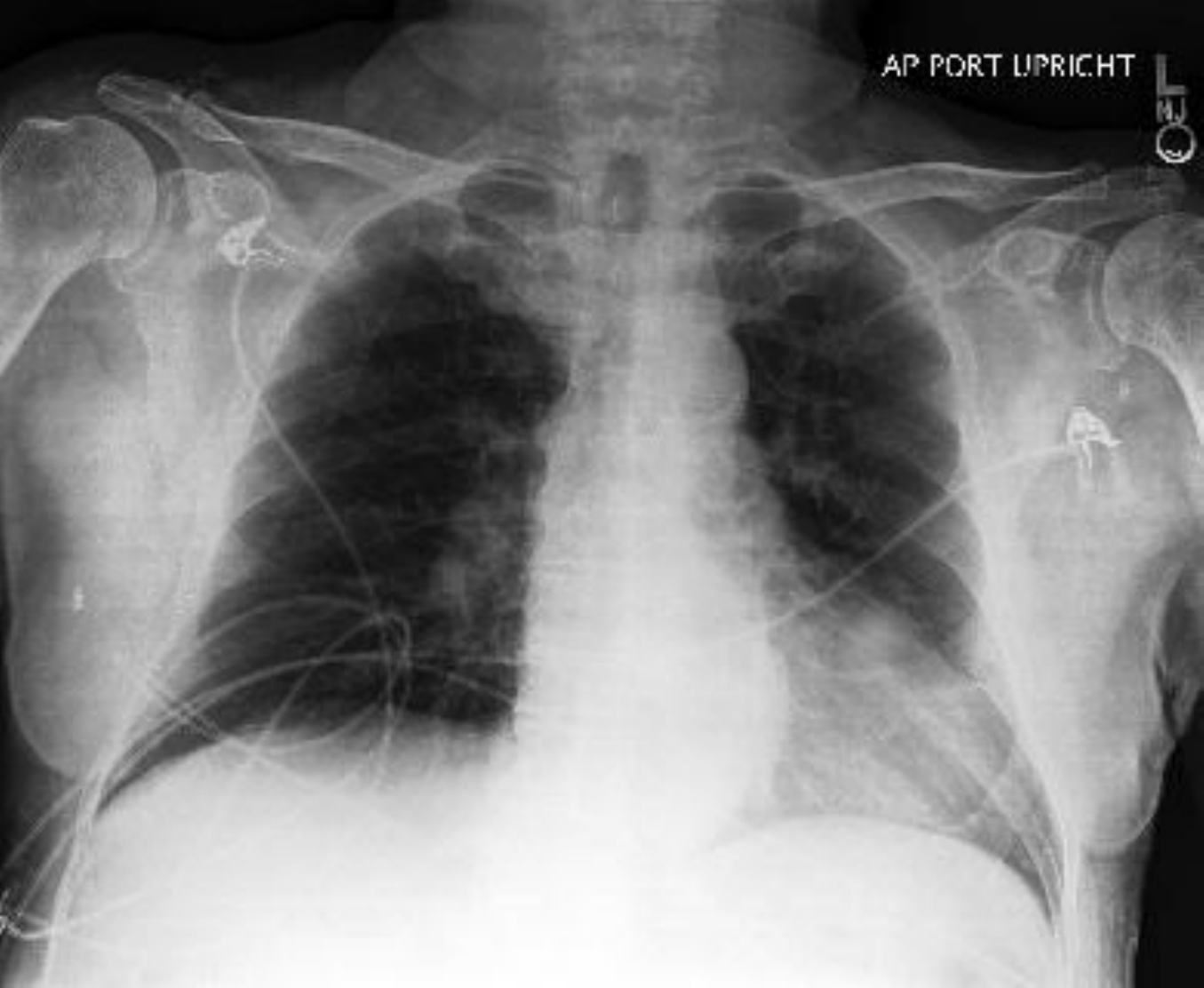} &   \includegraphics[width=30mm]{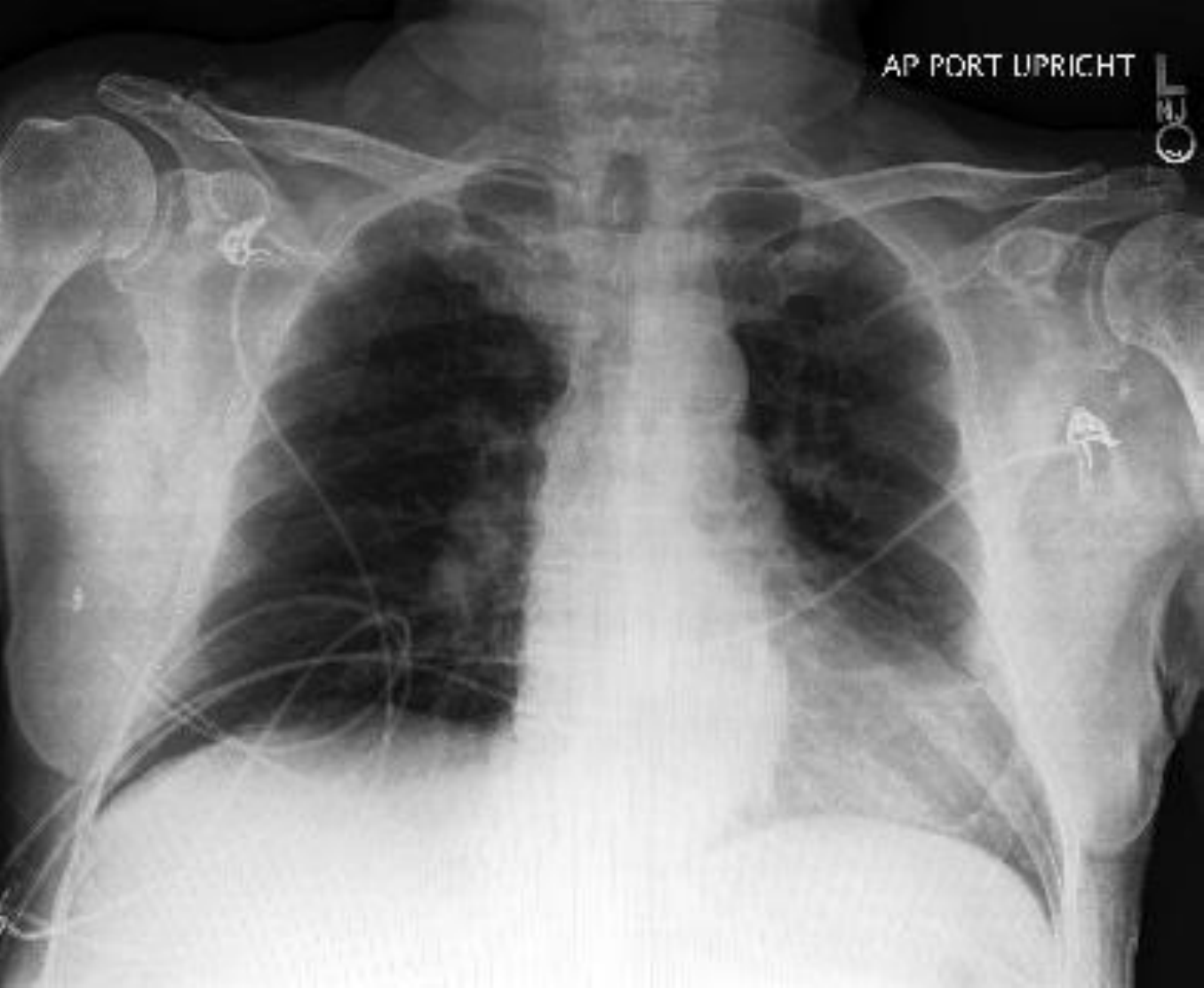} &
  \includegraphics[width=30mm]{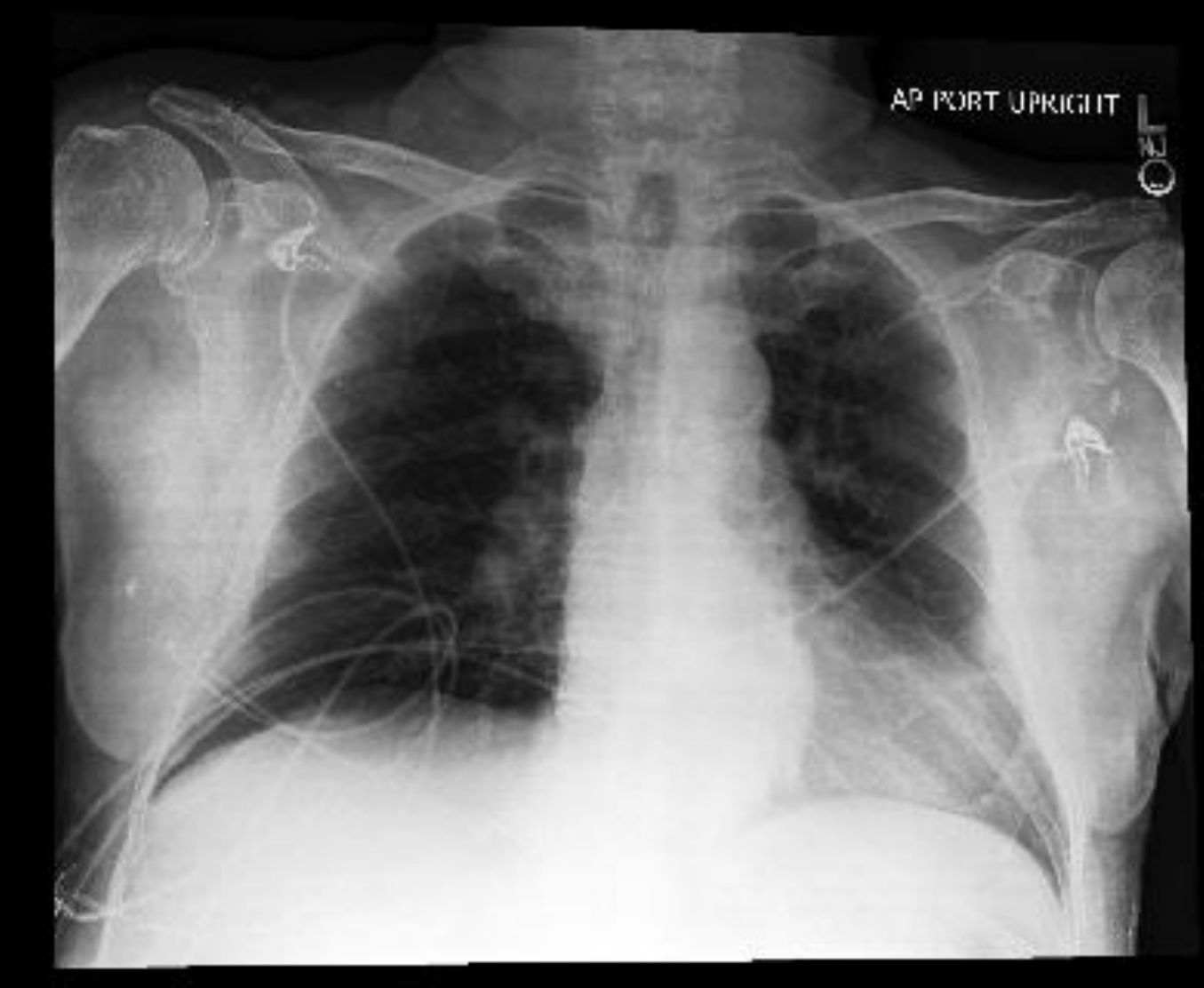} \\
(g) Glare Matte & (h) Moir\'e & (i) Tilt \\[6pt]
\end{tabular}
}

\end{table*}

Digital transformations were produced by successive random alterations of contrast and brightness. First, the image was either enhanced for greater or lesser contrast. Setting a contrast factor of 1 for the original image, the contrast up transformation increased contrast by a factor of 1.1 and the contrast down transformation decreased contrast by a factor of 0.83. For both these factors, random noise between -0.01 and 0.01 was applied. After the contrast modification, the brightness of the image was then transformed randomly up or down using the same numeric factors. Both the brightness and contrast transformations were applied using the Python PIL ImageEnhance class.

\begin{table*}[htbp]
	\begin{minipage}{0.5\linewidth}
		\floatconts
		{tab:imgcounts}
		{\caption{ The number of patients, studies, \\and images in CheXphoto.}}
    	{\begin{tabular}{llll}
                              Dataset       &  Patients     & Studies       &  Images \\
            \hline
                            \textit{Training} \\
                            iPhone & 295 & 829 & 1,000 \\
                            Nokia & 3,000 & 8,931 & 10,507 \\
                            Synthetic & 3,000 & 8,931 & 10,507 \\
                            \hline
                            \textit{Validation}  \\
                            Natural & 200 & 200 & 234 \\
                            Synthetic & 200 & 200 & 234 \\
                            Film & 200 & 200 & 200 \\
                            \hline
                            \textit{Test} & 500 & 500 & 668 \\
                            
            \hline
        \end{tabular}}
	\end{minipage}\hfill
	\begin{minipage}{0.5\linewidth}
		\floatconts{fig:Figure 3}
		{\captionof{figure}{CheXphoto directory structure}}
		{\includegraphics[width=65mm]{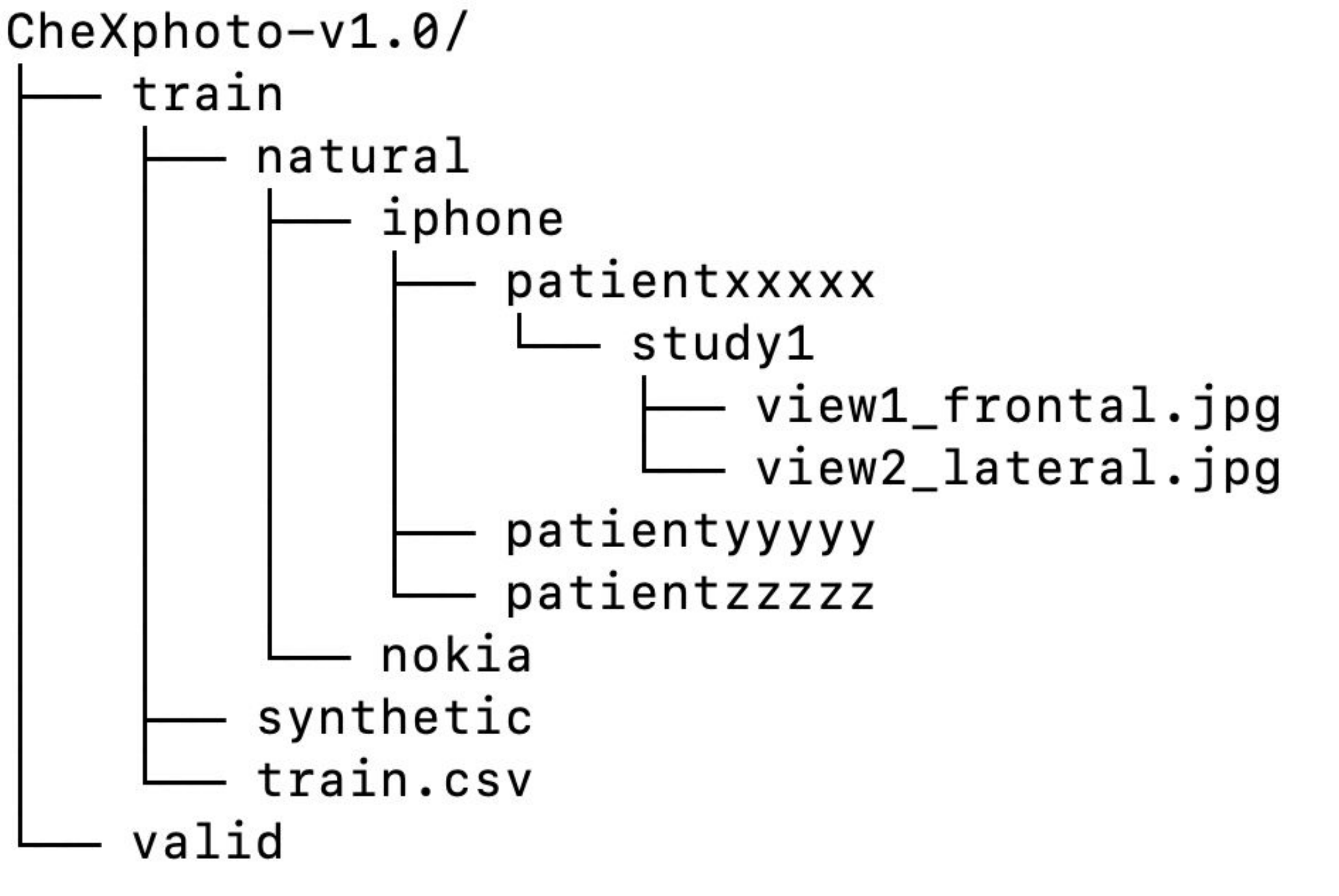}}
	\end{minipage}
\end{table*}

Spatial transformations consisted of alterations to add glare, moir\'e effects and perspective changes. First, we applied a glare matte transformation to simulate the effect of photographing a glossy film which reflects ambient light. This was produced by randomly generating a two-dimensional multivariate normal distribution which describes the location of a circular, white mask. Second, a moir\'e effect was added to simulate the pattern of parallel lines seen by digital cameras when taking pictures of computer screens. The effect is produced as a result of the difference in rates of shutter speed and LCD screen sweeping refresh rate. The moir\'e effect was simulated by generating semi-transparent parallel lines, warping them and overlaying them onto each image. Finally, a tilt effect was added to simulate random distortions of perspective that may arise in taking a photo at angle to the screen. The tilt effect was produced by randomly scaling the $x$ and $y$ values of each of the corners by a factor between 0 and 0.05 towards the center. This random movement of corners is used to skew the entire photo.

Both the digital and spatial transformations are provided in CheXphoto. Each transformation may be reproduced individually using the code provided. Additional transformations - glare glossy, blur, motion, rotation, translation - are also included.

\subsection{Validation and Test}

We developed a CheXphoto validation and test set to be used for model validation and evaluation. The validation set comprises natural photos and synthetic transformations of all 234 x-rays in the CheXpert validation set, and is included in the public release, while the test set comprises natural photos of all 668 x-rays in the CheXpert test set, and is withheld for evaluation purposes.

We generated the natural photos of the validation set by manually capturing images of x-rays displayed on a 2560 $\times$ 1080 monitor using a OnePlus 6 cell phone (16 megapixel camera with a Sony IMX 519 sensor), following a protocol that mirrored the iPhone1k dataset. Synthetic transformations of the validation images were produced using the same protocol as the synthetic training set. The test set was captured using an iPhone 8, following the same protocol as the iPhone1k dataset. 

The validation set contains an additional 200 cell phone photos of x-ray films for 200 unique patients. As photos of physical x-ray films, this component of the validation set is distinct from the previously described natural and synthetic transformations of digital x-rays. Films for 119 patients were sampled from the MIMIC-CXR dataset \citep{johnson2019mimiccxrjpg}, and films for 81 patients were provided by VinBrain, a subsidiary of Vingroup in Vietnam, and originally collected through joint research projects with leading lung hospitals in Vietnam. The film dataset spans 5 observation labels (atelectasis, cardiomegaly, consolidation, edema, pleural effusion), with 40 images supporting each observation. Observation labels for each image were manually verified by a physician. Images were captured using a VinSmart phone with a 12MP camera by positioning the physical x-ray film vertically on a light box in typical clinical lighting conditions, and images were automatically cropped and oriented. 

\subsection{Technical Validation}

CheXphoto was developed using images and labels from the CheXpert dataset \citep{chexpert}. Photography of x-rays was conducted in a controlled setting in accordance with the protocols documented in the Methods section, which were developed with physician consultation. Although CheXphoto contains multiple images for some patients, either from the same or different studies, there is no patient overlap between the training, validation, and test sets. Code developed for synthetic transformations is version controlled and made available as an open source resource for review and modification. All images are uniquely identifiable by patient ID, study ID, and view, and the age and sex of each patient is provided in the data description CSV file. The original, unaltered images can be obtained from the CheXpert dataset by the unique identifiers. 

The CheXphoto dataset is organized by by transformation; the training and validation sets contain directories corresponding to the method of data generation. Within each directory, the x-ray images are organized in subdirectories by a patient identifier, study ID, and one or more individual views. Images are stored as JPEG files, and image dimensions vary according to the method of generation. Each transformation set has an associated CSV file, which provides observation labels from the CheXpert dataset and relative paths to the corresponding images.

\subsection{Data Access}

The CheXphoto training and validation sets are available for download\footnote{\url{https://stanfordmlgroup.github.io/competitions/chexphoto}}. The CheXphoto test set is withheld for official evaluation of models. CheXphoto users may submit their executable code, which is then run on the private test set, preserving the integrity of the test results. The testing process is enabled by CodaLab \citep{codalab}, an online platform for collaborative and reproducible computational research. CodaLab Worksheets exposes a simple command-line interface, which enables users to submit a Docker image, dependencies, and the necessary commands to run their models. These features allow us to run arbitrary code submissions on the withheld test set. Once a user has successfully uploaded their code to CodaLab, we will evaluate their performance on the withheld test set and share results on a live leaderboard on the web. 

In addition, the code used to prepare the synthetically generated dataset is publicly available\footnote{\url{https://github.com/stanfordmlgroup/cheXphoto}}. The synthetic transformations can be reproduced by running the \texttt{synthesize.py} script with the appropriate \texttt{CSV} file containing the paths to the images for which the perturbation is to be applied. Detailed instructions on flags and usage are included in the repository README.

\section{Conclusion}
We believe that CheXphoto will enable greater access to automated chest x-ray interpretation algorithms worldwide, principally in healthcare systems that are presently excluded from the benefits of digital medicine. By facilitating the development, validation, and testing of automated chest x-ray interpretation algorithms with a ubiquitous technology such as smartphone photography, CheXphoto broadens access to interpretation algorithms in underdeveloped regions, where this technology is poised to have the greatest impact on the availability and quality of healthcare.

\bibliography{sample}

\end{document}